\documentclass[final]{aipproc}
\layoutstyle{6x9}

\usepackage{amsfonts}
\usepackage{amsmath}
\usepackage{amssymb}
\newcommand{\FP}{\mathcal J}
\newcommand{\ar}{\arrowvert}

\newcommand{\be}{\begin{equation}}
\newcommand{\ee}{\end{equation}}
\newcommand{\ba}{\begin{eqnarray}}
\newcommand{\ea}{\end{eqnarray}}

\newcommand{\bs}{\bf}
\begin{document}

\title{Coulomb gauge approach to scalar hadrons}


\author{Steve  Cotanch}{
  address={Department of Physics, North Carolina State University, 
 Raleigh NC  27695, USA}
}

\author{Ignacio General}{
  address={Bayer School of Natural  and
Environmental  Sciences,
Duquesne University, Pittsburgh, PA 15282, USA}
}

\author{Ping Wang}{
  address={Jefferson Laboratory, 12000 Jefferson Ave., Newport News, VA 23606, USA
}
}
\author{Felipe  Llanes-Estrada}{
  address={Depto. F\'isica Te\'orica I, Universidad Complutense de Madrid,
28040 Madrid, Spain}
}

\begin{abstract}
The Coulomb gauge model, involving an effective QCD Hamiltonian in the Coulomb gauge, is applied to scalar hadrons.  Mass predictions are
presented for both conventional $q {\bar q}$ meson and $q {\bar q}q {\bar q}$ tetra-quark states. Mixing matrix elements between these states
were also computed and diagonalized to provide a reasonable description of the scalar spectrum below 2 GeV.
 
\end{abstract}

\maketitle

\section{Introduction}

The scalar hadron spectrum has been a puzzling  problem that has attracted wide interest.  In particular for the isoscalar $J^{PC} =0^{+ +}$ channel, below 1 GeV the
$f_0(600)$ or $\sigma$ has had a somewhat confusing history while above 1 GeV the  existence of exotic glueballs, hybrid mesons
and tetra-quark systems has yet to be established.  In this paper we apply the Coulomb gauge (CG) model  to scalar systems to provide
further insight regarding their structure.  In the next few sections we discuss the model, summarize selected previous results and present
numerical predictions for the scalar spectrum.  More comprehensive details can be found in Refs. \cite{LC1,LChybrid,LC2,pomeron,LBC,Cotanch,Ignacio,Ignacio2}.

\section{Coulomb gauge model} 

In the Coulomb or transverse gauge the exact QCD Hamiltonian   is
\begin{eqnarray}
H_{\rm QCD} &=& H_q + H_g +H_{qg} + H_{C}    \\
H_q &=& \int d{\bs x} \Psi^\dagger ({\bs x}) [ -i
{\mbox{\boldmath$\alpha$\unboldmath}} \cdot
{\mbox{\boldmath$\nabla$\unboldmath}}
+  \beta m] \Psi ({\bs x})   \\
H_g &=& \frac{1}{2} \int\!\! d {\bs x} \!\! \left[ \FP^{-1}{ \boldmath
\Pi}^a({\bs x})\cdot \!\!  \FP {\boldmath
\Pi}^a({\bs x}) +{\bf B}^a({\bs x})\cdot{\bf B}^a({\bs x}) \right] \; \;  \; \; \\
H_{qg} &=&  g \int d {\bs x} \; {\bf J}^a ({\bs x})
\cdot {\bf A}^a({\bs x}) \label{eq:J.A}\\
H_C &=& -\frac{g^2}{2} \int d{\bs x} d{\bs y}\FP^{-1} \rho^a ({\bs x})
 K^{ab}( {\bs x},{\bs y}  ) \FP \rho^b ({\bs y})   \ ,
\end{eqnarray}
where  $\Psi$ is the quark field with current
quark mass $m$, $g$ is the QCD coupling constant, ${A}^a =({\bf A}^a, A^a_0)$ are the gluon fields satisfying the
Coulomb gauge condition, $\bs{\nabla}\cdot{\bf A}^a = 0$ $(a = 1, 2, ... 8)$,
 ${\boldmath \Pi}^a = -{ \bf E}^a_{tr} $ are the conjugate momenta and 
$
{\bf B}^a = \nabla \times {\bf A}^a + \frac{1}{2} g f^{abc} {\bf
A}^b \times {\bf A}^c 
$,
are the non-abelian chromodynamic fields.
The color densities, $\rho^a({\bs x}) =  \Psi^\dagger({\bs x}) T^a\Psi({\bs x})
+f^{abc}{\bf
A}^b({\bs x})\cdot{\boldmath \Pi}^c({\bs x})$, and quark currents,
${\bf J}^a =  \Psi^\dagger ({\bs x})
\mbox{\boldmath$\alpha$\unboldmath}T^a \Psi ({\bs x})$, contain the standard    $SU(3)$
color matrices,  $T^a = \frac{\lambda^a}{2}$, and structure constants, $f^{abc}$. The
Faddeev-Popov determinant, $\FP = {\rm det}(\mathcal M)$, is a measure of the gauge manifold curvature and
involves the color
matrix ${\mathcal M} = {\mbox{\boldmath$\nabla$\unboldmath}} \cdot
{\bf D}$ with covariant derivative ${\bf D}^{ab} =
\delta^{ab}{\mbox{\boldmath$\nabla$\unboldmath}}  - g f^{abc} {\bf
A}^c$. The kernel
in Eq. (5) is given by $K^{ab}({\bs x}, {\bs y}) = \langle{\bs x},
a|{\mathcal M}^{-1} \nabla^2 {\mathcal M}^{-1}  |{\bs y}, b\rangle$.
The Coulomb gauge Hamiltonian
preserves rotational invariance, 
 is renormalizable, permits resolution
of the Gribov problem, avoids
spurious retardation corrections, aids identification of dominant,
low energy  potentials and  introduces  only physical degrees of
freedom (no ghosts).

Our model entails two approximations,
replace the exact Coulomb kernel with a calculable confining potential and
use the lowest order, unit value for the the Faddeev-Popov determinant, giving
 the CG model
Hamiltonian, $H_{\rm CG} = H_q + H_g^{\rm CG} + H_{qg} + H_C^{\rm CG}  $
\begin{eqnarray}
H_g^{\rm CG} &=& \frac{1}{2} \int d {\bs x}\left[ {\boldmath
\Pi}^a({\bs x})\cdot {\boldmath \Pi}^a({\bs x}) +{\bf B}^a({\bs
x})\cdot{\bf B}^a({\bs x})
\right] \label{eq:non-abelian} \\
H_C^{\rm CG} &=& -\frac{1}{2} \int d{\bs x} d{\bs y} \rho^a ({\bs
x}) \hat{V}(\ar {\bs x}-{\bs y} \ar ) \rho^a ({\bs y})   \ .
 \label{model}
\end{eqnarray}
A Cornell type potential, ${\hat V}(r)=-{\alpha_s}/{r}+\sigma r$, 
is used for the confining kernel
with previously determined string tension, $\sigma=0.135$ GeV$^{2}$, and
$\alpha_s =0.4$.

Next, hadron states are expressed as dressed quark (anti-quark) Fock
operators, $B^{\dag}_{\lambda{\cal C}}$ ($
    D^{\dag}_{\lambda{\cal C}}$), with  helicity, $\lambda = \pm 1$, and color
${\cal C }= 1,2,3$ acting on the Bardeen-Cooper-Schrieffer (BCS) 
    model vacuum, $|\Omega \rangle$ (see Refs. \cite{LC2,Ignacio} for full details). 
The $q \bar{q}$ meson state is
\begin{eqnarray}
    |\Psi^{JPC}\rangle & = & \int \!\! \frac{d\bf{k}}{(2\pi)^3} \Phi^{JPC}_{\lambda_1 \lambda_2}({\bf{k}})
    B^{\dag}_{{\lambda_1}{\cal C}}({\bf{k}}) D^{\dag}_{{\lambda_2}{\cal C}}({\bf{-k}}) |\Omega \rangle \ .
  \end{eqnarray}
        For the tetra-quark system the
wave function ansatz 
  \begin{eqnarray} 
    |\Psi^{JPC}\rangle = \int \!\!
    \frac{d{\bf{q}}_1}{(2\pi)^3} \frac{d{\bf{q}}_2}{(2\pi)^3}
    \frac{d{\bf{q}}_3}{(2\pi)^3} \Phi^{JPC}_{\lambda_1 \lambda_2
    \lambda_3 \lambda_4}({\bf{q}}_1,{\bf{q}}_2,{\bf{q}}_3) \; \; \; \;
    \\ \nonumber 
    R^{{\cal C}_1{\cal C}_2}_{{\cal C}_3{\cal C}_4}
    B^{\dag}_{\lambda_1{\cal C}_1}({\bf{q}}_1)
    D^{\dag}_{\lambda_2{\cal C}_2}({\bf{q}}_2)
    B^{\dag}_{\lambda_3{\cal C}_3}({\bf{q}}_3)
    D^{\dag}_{\lambda_4{\cal C}_4}({\bf{q}}_4)|\Omega \rangle \ , ~~~
  \end{eqnarray}
  is adopted
 with  quark (anti-quark) $cm$ momenta
   ${{\bf{q}}_1}$, ${{\bf{q}}_3}$ (${{\bf{q}}_2}$, ${{\bf{q}}_4}$).
 The
expression for the matrix $R^{{\cal C}_1{\cal C}_2}_{{\cal C}_3{\cal
C}_4}$ depends on the specific color  scheme selected
\cite{Cotanch,Ignacio2}. For the color singlet-singlet scheme,
$[(3 \otimes \bar{3})_{1} \otimes (3 \otimes
\bar{3})_{1}]_1$, where the $q \bar{q}$ pairs, $A$ and $B$,  couple
to  color singlets,
$R^{{\cal C}_1{\cal
C}_2}_{{\cal C}_3{\cal C}_4}=\delta_{{\cal C}_1{\cal
C}_2}\delta_{{\cal C}_3{\cal C}_4}$.
This gives the lowest mass among 
the four  color representations. 
The spin  wave function part is, 
$\langle \frac{1}{2}  \frac{1}{2}  \lambda_1 \lambda_2 |s_A \lambda_A\rangle$
 $\langle  \frac{1}{2}  \frac{1}{2} \lambda_3 \lambda_4 |s_B \lambda_B\rangle$ 
$\langle s_A  s_B \lambda_A  \lambda_B |J 
\lambda_A+\lambda_B\rangle$,
a product of Clebsch-Gordan coefficients where
$J$ is the total angular momentum, ${\bs s}_A = {\bs s}_1 + {\bs s}_2$ and  ${\bs s}_B = {\bs s}_3 + {\bs s}_4$.  For all scalar tetra-quarks states the orbital angular momenta
are zero, consistent with the lowest energy state.
A Gaussian radial wavefunction is used
(see \cite{Ignacio2} for details)
$
f(q_A,q_B,q_I) =
e^{-\frac{q^2_A}{\alpha^2_A} - \frac{q^2_B}{\alpha^2_B} - \frac{q^2_I}{\alpha^2_I}} \ ,
$
with variational parameters $\alpha_A = \alpha_B$ and
$\alpha_I$
determined by minimizing the tetra-quark mass
\begin{eqnarray}
M_{J^{PC}} &=& {\langle\Psi^{JPC}|H_{\rm CG}|\Psi^{JPC}\rangle}  
= M_{self}+M_{qq}+M_{\bar{q}\bar{q}}+M_{q\bar{q}}+M_{annih} \ ,
\end{eqnarray}
which was  previously calculated \cite{Cotanch,Ignacio2}.
The respective
contributions are  the $q$ and $\bar{q}$ self-energy,
the $qq$, $\bar{q}\bar{q}$ and $q\bar{q}$ scattering, and the
$q\bar{q}$ annihilation.

\section{Application to scalar hadrons}

The most recent particle tabulation \cite{pdg}  lists nine isoscalar $0^{++}$ states:
$f_0(600),  f_0(980)$, $  f_0(1370), f_0(1500), f_0(1710), f_0(2020), f_0(2100), f_0(2200)$ and $f_0(2330)$, of
which the last four are not included in the summary table.  Our model has previously predicted that there is one $gg$ glueball
around 1700 MeV \cite{pomeron} and two hybrids:  a $n\bar{n}g$ at 2135 MeV, where $n\bar{n}=\frac1{\sqrt{2}}(u\bar{u}+d\bar{d})$,
and a $s\bar{s}g$ at 2140 MeV.  In the pure quark sector, we also calculated \cite{LC2} two conventional $q \bar{q}$ states which were orbital p-waves:
a $n\bar{n}$ at 848 MeV and a $s\bar{s}$ at 1297 MeV.  Most recently we have  predicted \cite{Ignacio2} several tetra-quark states:
two $n\bar{n} n\bar{n}$ at 1282  and 1418 MeV and two $n\bar{n} s\bar{s}$ at 1582 and 1718 MeV.  There will also be a $s\bar{s} s\bar{s}$
as well as radially excited $n\bar{n}$ and $s\bar{s}$
states near to above 2 GeV which we have not calculated.  There are three model corrections which should be included and will modify
this predicted spectrum. The first is chiral symmetry which will significantly lower one of our predicted $n\bar{n} n\bar{n}$ states corresponding
to the $\pi \pi$ quantum number channel (this is the 1418 MeV state).  This effect is discussed further below and will be incorporated 
in a future analysis.  The second is to perform a  resonance or scattering calculation to obtain the imaginary part of the pole position or width
of the state.  This will also slightly affect the real part of the pole position, or resonance mass.  We have begun such an analysis \cite{bclr}
and will report subsequent developments elsewhere.  The third effect is mixing which we now address.

We only treat $q \bar{q}$ meson and tetra-quark mixing since in calculating 
 quark-hybrid and quark-glueball mixing matrix elements with our model Hamiltonian, the former are perturbative, and thus expected weak, while the latter
entirely vanish (mixing must proceed via higher order intermediate states).  This  suggests that glueball widths might  not be large, as typically expected,   consistent with a recent theoretical prediction \cite{bclr}.  Mixing with gluonic states clearly merits  further  study
which we plan to address in a future analysis.
Using the notation, $|q \bar{q}>$ and $|q \bar{q} q \bar{q}>$ for $|\Psi^{JPC}>$, the  mixed state is  given by
$
|J^{PC}\rangle =
a|n\bar{n}\rangle+b|s\bar{s}\rangle+c_i|n\bar{n}n\bar{n}\rangle_i+d_i|n\bar{n}s\bar{s}\rangle_i
$ for $i = 1, 2$.
The coefficients $a,b,c_i$
and $d_i$ are determined by diagonalizing the Hamiltonian matrix, in which
the meson-tetra-quark off-diagonal mixing element is (only  $H_C^{\rm CG}$  contributes) 
$
M = {\langle q\bar{q}|H_C^{\rm CG}|q\bar{q}q\bar{q}\rangle} \ ,
$
where $| q\bar{q} \rangle$ is $| n\bar{n}\rangle$ or $|
s\bar{s}\rangle$, and $| q\bar{q}q\bar{q} \rangle$ is $|
n\bar{n}n\bar{n}\rangle$ or $| n\bar{n}s\bar{s}\rangle$.
There are 12 off-diagonal matrix elements however three,
$\langle s\bar{s}|H_C^{\rm CG}|n\bar{n}\rangle$ and $\langle s\bar{s}|H_C^{\rm CG}|n\bar{n}n\bar{n}\rangle_i$,
vanish and four, $_i\langle n \bar{n}  n\bar{n}|H^{\rm CG}_C|n\bar{n} s \bar{s}\rangle_i$, are numerically very small.
The remaining   mixing matrix
elements  are, $\langle n\bar{n}|H^{\rm CG}_C|n\bar{n}n\bar{n}\rangle_i$,
$\langle n\bar{n}|H_C^{\rm CG}|n\bar{n}s\bar{s}\rangle_i$ and $\langle
s\bar{s}|H_C^{\rm CG}|n\bar{n}s\bar{s}\rangle_i$. For our model Hamiltonian, there are two types  of
mixing diagrams illustrated in Fig.~1. Because
of color factors,  nonzero mixing only exists   for
$q \bar{q}$ annihilation   between  different singlet
$q \bar{q}$ clusters. The  first diagram gives
\be
 M_1 = \frac12 \int\!\!\! \; d{\bf
q}_1^{} d{\bf q}_2^{} d{\bf q}_3^{}V(k)
{\cal U}_{\lambda_1}^\dag({\bf q}_1^{})
{\cal U}_{\lambda'_1}(-{\bf q}_4) F_{\lambda_1  \lambda_4}({\bf q}_1, {\bf q}_2, {\bf q}_3)
\Phi_{\lambda'_1\lambda_4}^{JPC}(-2{\bf
q}_4) \ ,
\ee
with   ${\bf q}_4= -{\bf q}_1 - {\bf k}$,  ${\bf
k}={\bf q}_2+{\bf q}_3$,
$F_{\lambda_1  \lambda_4}({\bf q}_1, {\bf q}_2, {\bf q}_3) = {\cal U}_{\lambda_3^{}}^\dag({\bf q}_3^{})
{\cal V}_{\lambda_2^{}}({\bf q}_2^{})\Phi_{\lambda_1^{}
\lambda_2^{} \lambda_3^{}\lambda_4^{}}^{JPC\dag}({\bf
q}_1^{},{\bf q}_2,{\bf q}_3^{})$ and dressed, BCS spinors ${\cal U}_\lambda $
 and ${\cal V}_\lambda $.
 The
effective confining potential in momentum space is $V(k)$. 
The   second diagram  yields
\begin{eqnarray}
 M_2 = \frac12 \int\!\!\! \; d{\bf
q}_1 d{\bf q}_2 d{\bf q}_3 V(k)
{\cal V}_{\lambda_4}^\dag({\bf q}_4)
{\cal V}_{\lambda'_4}({-\bf q}_1) F_{\lambda_1 \lambda_4}({\bf q}_1, {\bf q}_2, {\bf q}_3) \Phi_{\lambda_1\lambda'_4}^{JPC}(2{\bf q}_1) \ .
\end{eqnarray}
\begin{figure}[tp]
\hspace{.5cm}
\includegraphics[scale=0.5]{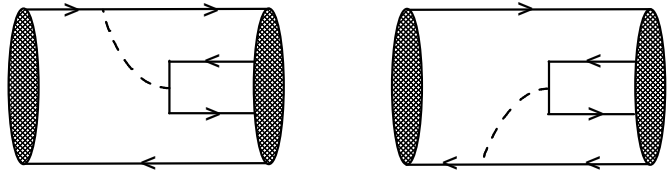}
\caption{Diagrams for the meson, tetra-quark mixing term.}
\end{figure}
The two Hamiltonian parameters in our model were independently determined
while the wavefunction parameters were obtained variationally.  Because we seek new model masses, the unmixed variational basis states
need not be ones producing a minimal, unmixed mass, so we selected one of the
variational parameters, $\alpha_I$, to provide an optimal mixing prediction and then
 studied the mixing sensitivity to
 this parameter.

\begin{figure}[b]
\includegraphics[scale=0.7]{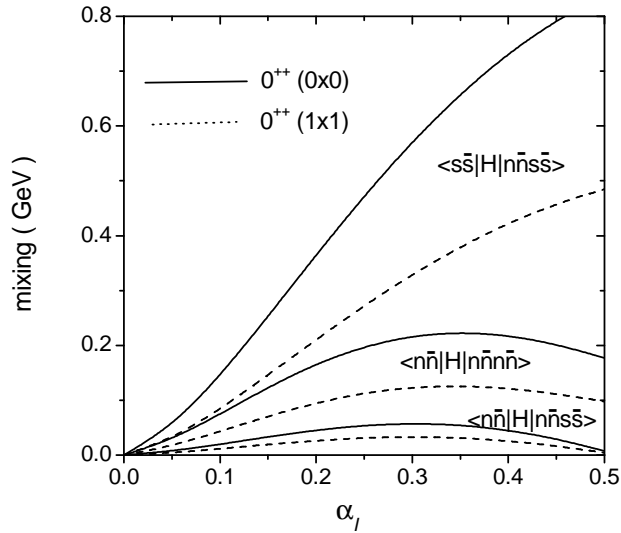}
\caption{Mixing matrix elements versus
$\alpha_I$. Solid (dashed) lines are for $q \bar{q}$ spin 0 (1).}
\end{figure}

For the $0^{++}$ tetra-quark state, the spin of the two $q \bar{q}$ clusters 
must either be $s_A=s_B=0 ~{\rm or}~ s_A=s_B=1$,
and for each the three mixing matrix
elements  versus $\alpha_I$ are shown  in Fig.~2.
The  mixing
term is zero when $\alpha_I$ is zero and then  increases with
increasing $\alpha_I$.
Note that mixing with $s \bar{s}$ states is stronger than with $n \bar{n}$ states.

Using the calculated matrix elements and  previously
predicted unmixed meson and tetra-quark masses \cite{LC2,Cotanch,Ignacio2},
 the complete Hamiltonian matrix was diagonalized to obtain the expansion  coefficients and masses
for the new eigenstates.   For   $\alpha_I=0.2$, the results for $0^{++}$ states are
compared in Table 1 to
 the  observed \cite{pdg} lowest six $0^{++}$ states. Noteworthy, after mixing, the
$n \bar{n}$ meson mass is shifted from 848 MeV to 783 MeV and the strange
scalar meson mass also decreases from 1297 MeV to 1026 MeV, now
close to the experimental value of 980 MeV. Mixing clearly improves the model predictions as the masses of the other
$f_0$ states are also in better agreement with data.  Figure 3 illustrates the over all
improved description that mixing provides for the $f_0$ spectrum.
New structure insight has also been obtained from the
coefficients, with the  predictions that the $\sigma / f_0(600)$ is predominantly a mixture 
of $n\bar{n}$ and $n\bar{n}n\bar{n}$ states while the $f_0(980)$
 consists mainly of $s\bar{s}$ and
$n\bar{n}s\bar{s}$ states.
\begin{table}[t]
    \caption{Mixing coefficients and masses in MeV for isoscalar $0^{++}$ states.}
    \begin{tabular}{ccccccc}
    \hline \hline \noalign{\smallskip}
   & $|n\bar{n}>$ & $|s\bar{s}>$ & $|n\bar{n}n\bar{n}>_1$ &
$|n\bar{n}n\bar{n}>_2$ & $|n\bar{n}s\bar{s}>_1$& $|n\bar{n}s\bar{s}>_2$ \\
    \hline \noalign{\smallskip}
no mixing & 848 & 1297 & 1282  & 1418 & 1582 & 1718  \\
mixing & 783 & 1026 & 1295 & 1466 & 1611 & 1962 \\
exp. & $f_0(600)$ & $f_0(980)$ & $f_0(1370)$ & $f_0(1500)$ & $
f_0(1710)$ & $f_0(2020)$ \\
& 400 - 1200 & $980 \pm 10$ & 1200 - 1500 & $1507 \pm 5$ & $ 1718\pm
2$
& $1992 \pm 16$ \\
    \hline \noalign{\smallskip}
coeff.  & a & b &  $c_1$ & $c_2$ & $d_1$ & $d_2$  \\
$f_0(600)$  & 0.947 & -0.070 & 0.138  & 0.277 & 0.006 & 0.030 \\
$f_0(980)$ & 0.052 & 0.838  & -0.012 & -0.021& 0.122 & 0.529 \\
$f_0(1370)$ & -0.181& 0.026  & 0.973  & 0.143 & 0.002 & 0.004 \\
$f_0(1500)$ & -0.248& 0.030  & -0.187 & 0.950 & 0.005 & 0.009 \\
$f_0(1710)$ & -0.036& -0.315 & 0.000 & -0.007 &0.901 & 0.295 \\
$f_0(2020)$ & -0.053& -0.439 & 0.000 & -0.006& -0.416 & 0.795 \\
    \hline \hline
    \end{tabular}
\end{table}
\begin{figure}[tp]
\includegraphics[scale=0.43]{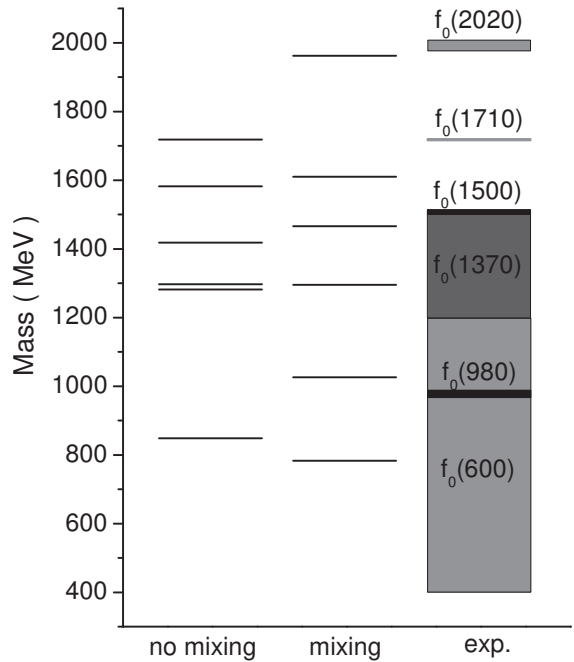}
\caption{Predicted unmixed, mixed and experimental $f_0$ spectrum.}
\end{figure}
There is general consensus  that the $\sigma / f_0(600)$ state
is a $\pi \pi$ resonance (pole in the $\pi \pi$ scattering amplitude) and thus has a
predominantly tetra-quark nature.
 Because the current numerical treatment of our CG model neglects chiral symmetry,
 our mass predictions for tetra-quark states which  couple to  $\pi \pi$ quantum numbers will generally be too
  heavy and should be regarded as preliminary.  A chiral symmetry preserving RPA
variational mixing study will be conducted in the future which should yield a lighter
scalar tetra-quark mass since our approach is formally equivalent to the Schwinger-Dyson treatment
which  previously predicted the $\sigma$ is a $\pi \pi$ resonance \cite{RPA}.


\section{Summary and conclusions}

Using the established CG model we have calculated $q\bar{q}$ and $q\bar{q}q\bar{q}$
mixing for the low-lying $0^{++}$ spectrum. 
We find mixing effects are important and necessary for an improved hadronic description.
Future work will  address mixing applications to
glueball and hybrid mesons to facilitate 
establishing their existence.


\begin{theacknowledgments}
  The authors wish to commend the organizers of Scadron70 for  a 
very productive workshop. Steve Cotanch also thanks George Rupp and Pedro Bicudo  for their
kind assistance and hospitality. Work supported by grants DOE DE-FG02-03ER41260, BSCH-PR34/07-15875, 
FPA 2004 02602, FPA 2005-02327
and Acci\'on Integrada Hispano-Portuguesa HP2006-0018.
\end{theacknowledgments}

\end{document}